\documentclass[aps,article,multicol,twocolumn,epsfig,showpacs]{revtex4}
\usepackage{graphicx}

\begin{document}

\preprint{HEP/123-qed}
\title{Cooperative behavior of qutrits with dipole-dipole interactions}
\author{A. Mandilara and V. M. Akulin}
\affiliation{Laboratoire Aim\'{e} Cotton, B\^{a}t. 505, CNRS II, Campus d'Orsay, ORSAY\
CEDEX F-91405, FRANCE}

\begin{abstract}
We have identified a class of many body problems with  analytic solution beyond the
mean-field approximation. This is the case where each body can be considered
as an element of an assembly of interacting particles that are
translationally frozen multi-level quantum systems and that do not change
significantly their initial quantum states during the evolution. In
contrast, the entangled collective state of the assembly experiences an
appreciable change. We apply this approach to interacting three-level
systems.
\end{abstract}

\startpage{1}

\maketitle

The description of the collective behavior of quantum ensembles beyond the
mean-field approximation is one of the most challenging tasks of modern
physics. The experiments with three-level Rydberg atoms performed during
last decade \cite{1,2,3} unambiguously point out the important role of
essentially many-body phenomena in frozen gases with interaction among the
internal degrees of freedom \cite{16}. Besides, the three-level systems (qutrits) have
been considered \cite{4} in the context of quantum informatics where, in
particular, the questions of generalized entanglement \cite{5} and coherence
protection \cite{6} have been addressed recently. Therefore, 
analytical results considering dynamics of quantum assemblies \cite{7} of
three-level elements beyond the mean-field approach and hence allowing for
the multipartite entanglement have become an issue of general interest.

Here we present the exhaustive description of an important particular class
of quantum states of an assembly of $N$ interacting identical qutrits where
each single qutrit state remains close to the initial state with
predominantly populated middle level. Still the collective assembly quantum
state, being essentially entangled, differs from a product state typical of
the mean-field approximation and moreover it can considerably deviate from the
initial state. Significant deviation of the collective state which occurs in
spite of small deviations of the single-particle states is known in the
many-body theory as the orthogonality catastrophe. Our approach can be
generalized to assemblies of arbitrary multilevel elements that  remain close
to their initial states.

For the description we employ a technique of nilpotentials inspired by the
ideas of the Glauber coherent states \cite{8}, which has been adapted for
three-level systems and employed for the description of quantum entanglement 
\cite{9}. The collective state under consideration 
\begin{equation}
\left\vert \Psi _{W}\right\rangle =\mathrm{e}^{%
\sum_{i,j=1}^{N}u_{j}^{+}W_{ij}t_{i}^{+}}\left\vert \mathsf{O}\right\rangle
\label{Eq.(1)}
\end{equation}%
can be represented in terms of the commuting operators $t_{i}^{+}$ and $%
u_{i}^{+}$ defined as the nilpotent $su(3)$ operators \cite{10} that create
the upper $\left\vert 1\right\rangle _{i}=t_{i}^{+}\left\vert 0\right\rangle
_{i}$ and the lower $\left\vert -1\right\rangle _{i}=u_{i}^{+}\left\vert
0\right\rangle _{i}$ states, respectively, by acting on the middle state $%
\left\vert 0\right\rangle _{i}$ of $i$-th qutrit. Here $\left\vert \mathsf{O}%
\right\rangle =\prod_{i}\left\vert 0\right\rangle _{i}$ denotes the initial
state of the assembly where all qutrits are in the middle state \cite{foot}. The
requirement\textit{\ (i)} that each qutrit is close to the initial state
implies $\left\vert W\right\vert =\mathrm{Tr}WW^{+}\ll N$. The assembly
state ~(\ref{Eq.(1)}) is not normalized to unity, although the
entanglement matrix $W_{ij}$ contains all information about the state,
including normalization. It is expedient to explicitly give the
normalization factor $\left\langle \Psi _{W}\right. \left\vert \Psi
_{W}\right\rangle $ and the population $n_{1}$ of the upper states of
qutrits $\left\vert 1\right\rangle _{i}$%
\begin{equation}
\left\langle \Psi _{W}\right. \left\vert \Psi _{W}\right\rangle =\exp \{%
\mathrm{Tr}WW^{+}\},\ n_{1}=\mathrm{Tr}\frac{WW^{+}}{1-WW^{+}}
\label{Eq.(1a)}
\end{equation}%
corresponding to the state vector eq.(\ref{Eq.(1)}). It is also worth
mentioning that the sum $f=\Sigma _{i,,j=1}^{N}u_{j}^{+}W_{ij}t_{i}^{+}$
represents the tanglemeter \cite{9} of the state $\left\vert \Psi
_{W}\right\rangle $. Due to our initial assumption $f$ is not of the
most generic form,i.e. it is lacking terms like $\Sigma _{i,,j=1}^{N}u_{j}^{+}A_{ij}u_{i}^{+}$ or
 $\Sigma _{i,,j=1}^{N}t_{j}^{+}B_{ij}t_{i}^{+}$. Nevertheless $f$ contains  cross terms which
  ensure, according to 
  the entanglement criterion \cite{9}, the existence of 
  entanglement among the qutrits of the assembly under consideration.

We now consider an assembly of qutrits with dipole-dipole interaction $%
\sum_{i,<j}V_{ji}\widehat{d}_{j}\widehat{d}_{i}$ and find the time-dependent
matrix $\widehat{W}$. For the dipole moment operator $%
d_{i}=u_{i}^{+}+t_{i}^{-}+u_{i}^{-}+t_{i}^{+}$ the interaction Hamiltonian
reads%
\begin{equation}
\widehat{H}_{int}=\sum\limits_{i\neq
,j}(u_{j}^{+}+t_{j}^{-})V_{ji}(u_{i}^{-}+t_{i}^{+}).  \label{Eq.(2)}
\end{equation}%
We assume that the single qutrit Hamiltonian has the form%
\begin{equation}
\widehat{H}_{0}=\beta \lambda ^{(3)}+\alpha \frac{\lambda ^{(8)}}{\sqrt{3}}%
=\left( 
\begin{array}{ccc}
\beta +\frac{\alpha }{3} & 0 & 0 \\ 
0 & -\frac{2\alpha }{3} & 0 \\ 
0 & 0 & -\beta +\frac{\alpha }{3}%
\end{array}%
\right) ,
\end{equation}
and that the coupling matrix $V_{ji}$ has a spectral decomposition $%
V_{ji}=\sum_{m}C_{jm}V_{m}C_{mi}$. \ We also assume that $\beta \gg \alpha $
and exclude the high frequency $\beta $ in the rotating wave approximation.
Then the evolution operator in the interaction representation can be written
as a functional integral%
\begin{equation}
\exp \left\{ -\mathrm{i}\widehat{H}t\right\} =\frac{1}{A}\int \mathrm{e}^{-%
\mathrm{i}S}\prod_{m}\mathcal{D}Z_{m}(t)\mathcal{D}Z_{m}^{\ast }(t),
\label{Eq.(3)}
\end{equation}%
where the normalization constant $A$ and the action%
\begin{eqnarray*}
S &=&\int \left[ \sum_{m}Z_{m}(t)Z_{m}^{\ast
}(t)-\sum_{m,j}V_{m}^{1/2}C_{jm}(\mathrm{e}^{\mathrm{i}\alpha
t}u_{j}^{+}\right.  \\
&&+\mathrm{e}^{-\mathrm{i}\alpha t}t_{j}^{-})Z_{m}(t) \\
&&\left. -\sum_{m,j}V_{m}^{1/2}C_{mj}(\mathrm{e}^{-\mathrm{i}\alpha
t}u_{j}^{-}+\mathrm{e}^{\mathrm{i}\alpha t}t_{j}^{+})Z_{m}^{\ast }(t)\right] \mathrm{d}t
\end{eqnarray*}%
are given in terms of the components $Z_{m}(t)$ and $Z_{m}^{\ast }(t)$ of
two complex conjugated vector-function variables.

The representation ~(\ref{Eq.(3)}) allows one to consider dynamics of
qutrits independently: each qutrit now is subject to an action of the
single-particle time-dependent Hamiltonian%
\begin{equation}
\widehat{H}_{i}\left( t\right) =(\mathrm{e}^{\mathrm{i}\alpha t}u_{i}^{+}+%
\mathrm{e}^{-\mathrm{i}\alpha t}t_{i}^{-})E_{i}(t)+(\mathrm{e}^{-\mathrm{i}%
\alpha t}u_{i}^{-}+\mathrm{e}^{\mathrm{i}\alpha t}t_{i}^{+})E_{i}^{\ast }(t)
\label{Eq.(4)}
\end{equation}%
where%
\begin{equation}
E_{i}(t)=\sum_{m}V_{m}^{1/2}C_{im}Z_{m}(t)  \label{Eq.(5)}
\end{equation}%
is an \textquotedblleft effective electric field\textquotedblright\
depending on the functional variables $Z_{m}(t)$. \ The requirement \textit{%
(i)} justifies the employment of the second-order time-dependent
perturbation theory expression 
\begin{equation}
\widehat{U}_{i}(t)\simeq 1-\mathrm{i}\int^{t}\widehat{H}_{i}\left( x\right) 
\mathrm{d}x-\int^{t}\widehat{H}_{i}\left( y\right) \int^{y}\widehat{H}%
_{i}\left( x\right) \mathrm{d}x\mathrm{d}y
\end{equation}
for the evolution operator of $i$-th qutrit initially in the state $%
\left\vert 0\right\rangle _{i}$, which we write down in the form 
\begin{eqnarray}
\widehat{U}_{i}(t)\left\vert 0\right\rangle _{i} &\simeq &\mathrm{e}^{-%
\mathrm{i}u_{i}^{+}\int^{t}\mathrm{e}^{\mathrm{i}\alpha x}E_{i}(x)\mathrm{d}%
x-\mathrm{i}t_{i}^{+}\int^{t}\mathrm{e}^{\mathrm{i}\alpha x}E_{i}^{\ast }(x)%
\mathrm{d}x}  \nonumber \\
&&\mathrm{e}^{-\int^{t}E_{i}(y)\int^{y}\mathrm{e}^{\mathrm{i}\alpha
(y-x)}E_{i}^{\ast }(x)\mathrm{d}x\mathrm{d}y}  \label{Eq.(5a)} \\
&&\mathrm{e}^{-\int^{t}E_{i}^{\ast }(y)\int^{y}\mathrm{e}^{\mathrm{i}\alpha
(y-x)}E_{i}(x)\mathrm{d}x\mathrm{d}y}\left\vert 0\right\rangle _{i}\ . 
\nonumber
\end{eqnarray}%
It is due to this very approximation that the problem becomes analytically
soluble.

Substitution of ~(\ref{Eq.(5a)}) into ~(\ref{Eq.(3)}) with the allowance
for the relation $\sum_{i}V_{k}^{1/2}V_{m}^{1/2}C_{im}C_{ik}=\delta
_{km}V_{m}$ yields the action%
\begin{eqnarray}
S &=&\sum_{m}\int Z_{m}(t)Z_{m}^{\ast }(t)\mathrm{d}t  \label{Eq.(7)} \\
&&-\sum_{j,m}V_{m}^{1/2}C_{jm}\int^{t}\mathrm{e}^{\mathrm{i}\alpha x}\left(
u_{j}^{+}Z_{m}(x)+t_{j}^{+}Z_{m}^{\ast }(x)\right) \mathrm{d}x  \nonumber \\
&&-\sum_{m}V_{m}\int^{t}\int^{y}\mathrm{e}^{\mathrm{i}\alpha
(y-x)}Z_{m}(y)Z_{m}^{\ast }(x)\mathrm{d}x\mathrm{d}y  \nonumber \\
&&-\sum_{m}V_{m}\int^{t}\int^{y}\mathrm{e}^{\mathrm{i}\alpha
(y-x)}Z_{m}^{\ast }(y)Z_{m}(x)\mathrm{d}x\mathrm{d}y  \nonumber
\end{eqnarray}%
which is bilinear in the field variables $Z_{m}$ and $Z_{m}^{\ast }$. This
allows one to exactly evaluate the Gaussian functional integral by
performing standard calculations: from the action $S$ of ~(\ref{Eq.(7)})
one derives the Lagrange equations $\frac{\delta S}{\delta Z_{m}(x)}=0$, $%
\frac{\delta S}{\delta Z_{m}^{\ast }(x)}=0$ for the extremum trajectory. By
substituting the solutions of these equations 
\begin{eqnarray}
Z_{m}(x) &=&-\frac{\left( \mathrm{i}\alpha \sin x\omega _{m}+\omega _{m}\cos
x\omega _{m}\right) \sum_{i}V_{m}^{1/2}C_{mi}t_{i}^{+}}{\mathrm{i}\left(
\alpha +V_{m}\right) \sin t\omega _{m}+\omega _{m}\cos t\omega _{m}} 
\nonumber \\
Z_{m}^{\ast }(x) &=&\frac{\left( \mathrm{i}\alpha \sin x\omega _{m}+\omega
_{m}\cos x\omega _{m}\right) \sum_{i}V_{m}^{1/2}C_{mi}u_{i}^{+}}{\mathrm{i}%
\left( \alpha +V_{m}\right) \sin t\omega _{m}+\omega _{m}\cos t\omega _{m}} 
\nonumber
\end{eqnarray}%
with $\omega _{m}=\sqrt{\alpha \left( \alpha +2V_{m}\right) }$, to ~(\ref%
{Eq.(7)}) we find the part of $S$ which depends only on the operators $%
u_{i}^{+}$ and $t_{i}^{+}$ but not on the fields $Z_{m}$ and $Z_{m}^{\ast }$%
. The remaining part depending on the fields but not on $u_{i}^{+}$ and $%
t_{i}^{+}$ is a functional integral which gives a $c$-number and thus can be
ignored. The evaluation yields $\ \exp \left\{ -\mathrm{i}\widehat{H}%
t\right\} \left\vert \mathsf{O}\right\rangle \sim \exp \left\{
\sum\limits_{i,,j=1}^{N}u_{j}^{+}W_{ij}t_{i}^{+}\right\} \left\vert \mathsf{O%
}\right\rangle $ with the tanglemeter matrix%
\begin{equation}
\widehat{W}(t)=\frac{\widehat{V}}{\mathrm{i}\sqrt{\alpha (2\widehat{V}%
+\alpha )}\cot \left[ t\sqrt{\alpha (2\widehat{V}+\alpha )}\right] -\widehat{%
V}-\alpha }\ .  \label{Eq.(9)}
\end{equation}%
Substitution of ~(\ref{Eq.(9)}) to ~(\ref{Eq.(1a)}) yields%
\begin{equation}
n_{1}=\mathrm{Tr}\frac{\widehat{V}^{2}\sin ^{2}\left[ t\sqrt{\alpha (2%
\widehat{V}+\alpha )}\right] }{\alpha (2\widehat{V}+\alpha )}
\label{Eq.(10)}
\end{equation}

\begin{figure}
\begin{center}
{\centering{\includegraphics*[width=0.4\textwidth]{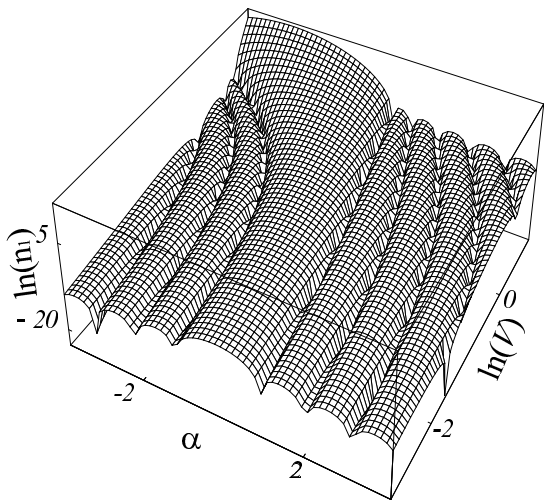}}} \end{center} \vspace{0.1cm}
\caption{Number $n_{1}$ of the qutrits in the upper state as a function of
the coupling $v=NV$ and the detuning $\protect\alpha $ for $t=3$.}
\label{Fig.1}
\end{figure}

We are now in the position to consider several particular examples. We start
with the case where all qutrits interact via a collective dipole moment,
that is when $V_{i,j}=V=\mathrm{const}$. In this case the matrix $\widehat{W}
$ has only one nonzero eigen value $V_{1}=NV$. This results in the upper
state population 
\begin{equation}
n_{1}=\frac{N^{2}V^{2}}{\alpha (2NV+\alpha )}\sin ^{2}\left[ t\sqrt{\alpha
(2NV+\alpha )}\right] \ ,  \label{Eq.(11)}
\end{equation}%
shown in figure \ref{Fig.1} as a function of the collective coupling $NV$ and
the detuning $\alpha $. The maximum rate of the creation of the qutrits in
the upper states corresponds to $\alpha =-NV$ where $n_{1}=\sinh ^{2}\left[
NVt\right] $. This expression is valid as long as $\sinh ^{2}\left[ NVt%
\right] \ll N$, \ that is when the assumption \textit{(i) }is fulfilled. 
One gets a deeper insight into the physical meaning of this result by
comparing the assembly of collectively interacting qutrits with the lasing
of an inverted two-level media \cite{11}. In this case the operator $t^{+}$
corresponds to the photon creation operator $a\dag $ while $u^{+}$
corresponds to the excitation annihilation operator $\sigma ^{-}$. As long
as the total number of the emitted photons remains much smaller than the
total number of the two-level atoms these two systems are almost equivalent,
-- the main difference being that the collective dipole-dipole interaction
shifts the resonance of the lasing rate from $\alpha =0$ to the point $%
\alpha =-NV$.

\begin{figure}
\begin{center}
{\centering{\includegraphics*[width=0.4\textwidth]{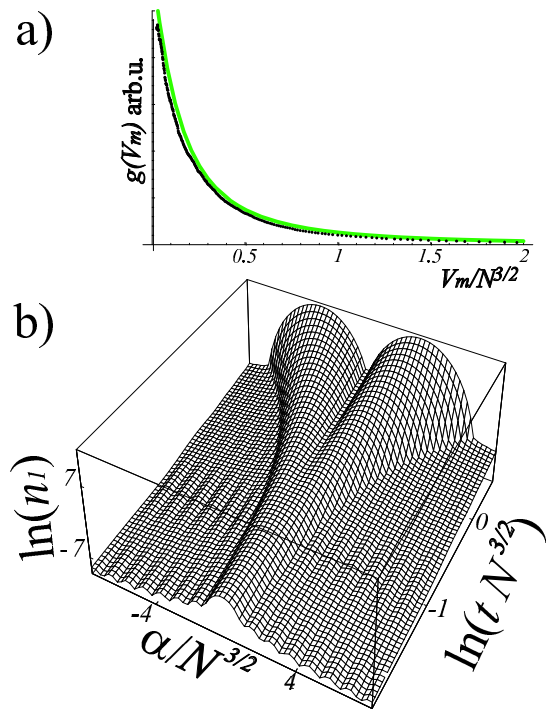}}} \end{center}\vspace{0.1cm}
\caption{a) Density of the eigenvalues $V_{m}$ for $N=300$ dipoles averaged
over $100$ random distributions in a unit cube (dots) and a heuristic fit $%
g(V_{m})=N/4V_{m}e^{\protect\sqrt{\protect\pi /2}}\cosh \frac{\ln 4V_{m}}{%
\protect\sqrt{\protect\pi }}$ (solid green line). We set the dipole moment $%
\protect\mu =1$. The distribution is symmetric with respect to zero. b)
Number $n_{1}$ of the qutrits in the upper state as a function of the scaled
time $tN^{3/2}$ and the scaled detuning $\protect\alpha /N^{3/2}$.}
\label{Fig2}
\end{figure}

Next example is an individual dipole-dipole coupling of qutrits with $%
V_{ji}=\mu ^{2}(1-2\cos ^{2}\theta _{ji})/r_{ji}^{3}$, where $r_{ji}$ is the
distance between $i$-th and $j$-th qutrits, $\theta _{ji}$ is the angle
between the radius-vector $\overrightarrow{r}_{ji}$ and the $z$ direction,
and $\mu $ is the dipole moment matrix element. We consider the position of
each qutrit as an independent random variable and assume that the qutrits
have uniform spatial density $n$. The statistical properties of the random
matrix corresponding to $1/r^{3}$ interaction in disordered media is a
challenging problem which has already been addressed in the context of spin
glasses \cite{12,13} and cold Rydberg atoms \cite{2, 4}. In particular, the
distribution $g(V_{m})$ of the eigenvalues of $V_{ij}$ found numerically 
\cite{14} has essentially non-analytical behavior near $V_{m}\rightarrow 0$,
in contrast with the well-known Wigner semicircular distribution of the
Gaussian random matrix eigenvalues. In figure \ref{Fig2} along with the results
of a similar numerical work performed for a larger statistical ensemble and
followed by a more accurate analytical fit to the distribution of $V_{m}$ we
depict the population ~(\ref{Eq.(10)}) averaged over this distribution $%
g(V_{m})$. \bigskip 

Note that the numerical results with $\mu =1$ are given in heuristic
dimensionless units $\alpha /N^{3/2}$. Yet unknown is the energy parameter
correctly describing the cooperative phenomena in ensembles with $1/r^{3}$
interaction. We guess that it might resemble the combination of parameters $%
\mu ^{2}n\sqrt{N}$ where $\mu ^{2}n$ is the typical two-particle interaction
while the typical parameter $\sqrt{N}$ allows for the cooperative effects
resembling Dicke superadience \cite{15} of two-level particles. However,
unambiguous identification of this parameter is the subject of a more
detailed future consideration. We also note that by assuming small
deviations of \ the qutrits from their initial state, we discard from the
consideration the strongly interacting dimers of qutrits that give an
important contribution to $n_{1}$ \cite{2} at large $\alpha $. The latter
effect requires a more sophisticated model, which would consider these
dimers as the assembly elements of another type.

The third example concerns a controlled behavior of the assembly where, as
earlier, the qutrits interact via the collective dipole, but variation of $%
\alpha $ is now possible during this interaction. This example shows that
the behavior of the three-level systems, though similar, still is richer
than that of the two-level systems and may display effects similar to mode
beats. We consider the case where $\alpha $, initially different from zero,
can be switched off at $t=t_{1}$, then remains zero during a time interval $%
\Delta t=t_{2}-t_{1}$, and takes the initial value for $t>t_{2}$. We notice
the change of the symmetry associated with such a control: a three-level
quantum system, which generically has the $su(3)$ symmetry, at $\alpha =0$
becomes equivalent to a spin-$1$ particle possessing the $su(2)$ symmetry.
For the description, in the equation (\ref{Eq.(7)}) for the action one has
to replace $\mathrm{e}^{\mathrm{i}\alpha x}$ and $\mathrm{e}^{\mathrm{i}%
\alpha (y-x)}$ by $\mathrm{e}^{\mathrm{i}\int_{0}^{x}\alpha (y)\mathrm{d}y}$
and $\mathrm{e}^{\mathrm{i}\int_{x}^{y}\alpha (s)\mathrm{d}s}$,
respectively. 
\begin{figure}
\begin{center}
{\centering{\includegraphics*[width=0.4\textwidth]{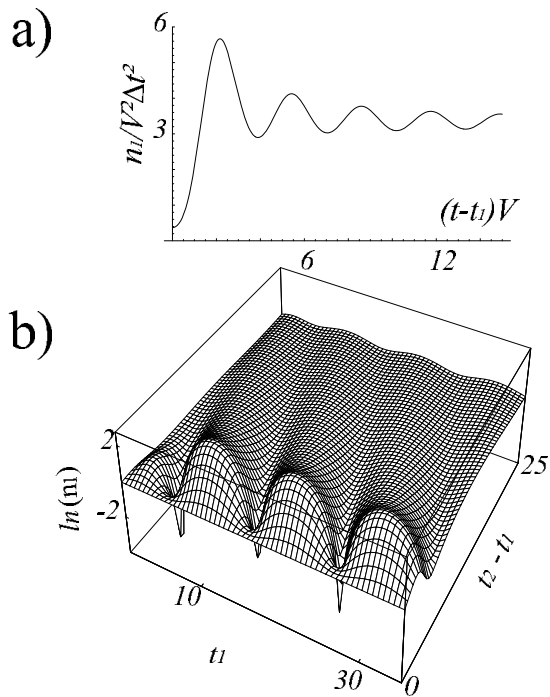}}}\end{center} \vspace{0.1cm}
\caption{(a) Number of the particles in the upper state $n_{1}$ for time
dependent $\protect\alpha $. We set $\protect\alpha =0$ when $t_{1}<t<t_{2}$
and the typical coupling $V\simeq \protect\mu ^{2}n\protect\sqrt{N}$. (b)
Ratio $n_{1}\left( NV,\protect\alpha ,t,t_{1},t_{2}\right) /n_{1}\left( NV,%
\protect\alpha ,t,t_{1},t_{1}\right) $ of the numbers of the qutrits in the
upper state as a function of the scaled time $t_{1}$ and the time difference 
$t_{2}-t_{1}$. We set $t=t_{2}+0.9$; $NV=-\protect\alpha =0.3$.}
\label{Fig3}
\end{figure}

When $V\ll \alpha $, the solution of the integral Langrange equations
results in 
\begin{equation}
n_{1}\simeq \frac{V^{2}}{\alpha ^{2}}\sin ^{2}\left[ \left( V+\alpha \right)
\left( t-\Delta t\right) \right]   \label{Eq.(13)}
\end{equation}%
for $\alpha \Delta t\ll 1$ and in 
\begin{equation}
n_{1}\simeq V^{2}\Delta t^{2}\left\{ 1+2\left[ 1-\cos V\left( t-t_{2}\right) %
\right] \left[ 1+\cos \alpha \left( t_{1}+t_{2}\right) \right] \right\} 
\label{Eq.(14)}
\end{equation}%
for $\alpha \Delta t\gg 1$. One can \ eliminate the second brackets in ~(%
\ref{Eq.(14)}) by averaging out the rapid oscillations at the frequency $%
\alpha $. \ In figure \ref{Fig3}(a) we depict $n_{1}\left( t-t_{2}\right) $
averaged not only over these oscillations but also over the distribution of
the collective couplings $g(V_{m})$ presented in figure \ref{Fig2}. One can see
that the population displays dying oscillations as a function of the
variable $t-t_{2}$ that is the time elapsed after the moment $t_{2}$ when
the detuning $\alpha $ is switched back. These oscillations can be
interpreted as beats of the symmetric $\frac{\left\vert -1\right\rangle
+\left\vert 1\right\rangle }{\sqrt{2}}$ and antisymmetric $\frac{\left\vert
-1\right\rangle -\left\vert 1\right\rangle }{\sqrt{2}}$ combinations of the
upper and the lower states of qutrits in the interaction representation. As
a direct consequence of $su(2)$ symmetry, the combination  $\frac{\left\vert
-1\right\rangle +\left\vert 1\right\rangle }{\sqrt{2}}$ evolves while $\frac{%
\left\vert -1\right\rangle -\left\vert 1\right\rangle }{\sqrt{2}}$ conserves
when $\alpha =0$. The  analytical solution is straightforward-- though
cumbersome, and it does not contain any principal technical complications.
In figure \ref{Fig3}(b) we depict the ratio $n_{1}\left( NV,\alpha
,t,t_{1},t_{2}\right) /n_{1}\left( NV,\alpha ,t,t_{1},t_{1}\right) $ as a
function of $t_{1}$ and $t_{2}$ obtained explicitly for $V=\mathrm{const}$
with the help of Mathematica package. Again one can see the population \
oscillations resulting from the quantum beats.

We conclude by summarizing the results obtained. \textit{(i) }The collective
behavior of an assembly of three-level elements (qutrits) admits an 
 exhaustive analytical description when each qutrit does not change
significantly its initial quantum state. This description is based on the
elements' dynamic separation with the help of a functional integral and
invokes \ the time-dependent second-order perturbation theory followed by
the exact evaluation of a Gaussian functional integral. \textit{(ii)} In the
simplest case where the interaction among the qutrits occurs via collective
dipole-dipole interaction, the dynamics of the systems resembles \ the
lasing of inverted two-level \ media: for the qutrits initially in the
middle state the collective downward transitions are accompanied by the
collective upward transitions \ similar to the photon creation in a coherent
state. The only difference is that the collective dipole-dipole interaction
shifts the resonance center such that the maximum transition rate occurs
when the frequencies of the downward and upward transitions differ by the
amount of the dipole collective coupling. \textit{(iii)} This effect
persists for the regular dipole-dipole interaction and yields a two-hump
frequency dependence of the transition rate. Positions of the maxima are
found with the help of the collective coupling distribution, which has been
numerically obtained for the media with $1/r^{3}$ interaction. \textit{(iv)}
Still three-level systems may have a more complex behavior than two-level
ones displaying quantum beats. This effect occurs when in the course of time
one changes the position of the middle level relative to the positions of
the upper and the lower levels switching in this way between $\ $the $su(2)$
and $\ su(3)$ symmetry of the qutrits.

A. M. acknowledges Ile de France for financial support. Both authors thank
Lorenza Viola and  V. A. is grateful to T. F. Gallagher, L. Ioffe, P. Pillet for stimulating discussions.


\begin{thebibliography}{0}
\bibitem{7} We employ the term \textquotedblleft assembly\textquotedblright\
for a collection of elements in fixed positions and distinguish it from an
ensemble, which implies an average over multiple possible realizations.
\bibitem{foot}  Eq.(\ref{Eq.(1)}) contains all information about the entanglement which is created in the ensemble due to the interaction, and that the result can be easily generalized on the case of an arbitrary initial product state.
\bibitem{10} One can consult reference \cite{9} for details.
\end{thebibliography}
\end{document}